\documentstyle[12pt,aaspp4,tighten]{article}

\begin{document}

\title{Where Are The Young Pulsars ?}
\author{Arnon Dar}
\affil{Department of Physics and Space Research Institute\\ 
   Technion, Haifa 32000, Israel}

\begin{abstract} 
We show that young pulsars with normal magnetic fields,
which are born as fast rotating neutron stars (NSs) in Type II/Ib
supernova explosions, can slow down quickly by relativistic particles
emission along their magnetic axis. When they slow down sufficiently, they
can undergo phase transitions and collapse to strange stars, quark stars or
black holes in explosions that remove a substantial fraction of their
initial angular momentum. The newly born strange or quark stars continue
to cool and spin down very quickly as soft gamma ray repeaters (SGRs) and
anomalous X-ray pulsars (AXPs) to become slowly rotating, radio-quiet, dim
X-ray pulsars or stellar black holes, depending on their initial mass. It
can explain the apparent dearth of young radio pulsars in SNRs and
in the Galaxy. 

\end{abstract} 
\section{Introduction}      

It is widely believed that neutron stars (NSs) which are born in Type
II/Ib supernova explosions (SNeII) of massive stars are rapidly rotating
neutron stars (NSs) with large magnetic fields, that are spinning down by
magnetic dipole radiation of their rotational energy (e.g. Shapiro and
Teukolsky 1983). Such Crab-like pulsars and their accompanying supernova
remnants (SNRs) are expected to be highly visible for tens of thousands of
years. The mean rate of SNeII in our galaxy has been estimated to be
${\rm R\approx 1/50 ~y}$ (Tamman and Van den Bergh 1991). It is consistent
with the 220 known young Galactic SNRs (Green 1998) if, on the average,
they are observable for $\sim 10^4$ y. Less than 30\% of the SNRs are Type
I SNe remnants (Tamman and Van den Bergh 1991) which do not produce NSs. 
But, despite detailed radio searches, only $<20$ radio pulsars out of over
1100 detected radio pulsars (Camilo et al.  1999) were found within/near
young SNRs and some of them can be due to chance overlap (Gaensler and
Johnston 1995). So where are all the young pulsars born in Galactic SNRs?

Traditional arguments invoking beaming, large natal kick velocities and
direct collapse to black holes have become less compelling with the
accumulating results from new surveys. Thus, an alternative explanation
for the apparent dearth of radio pulsars seems to be needed. Gotthelf and
Vasisht (1999) have suggested that newly born Crab-like NSs with a typical
surface magnetic field, ${\rm B \sim 10^{13}~Gauss}$, are rare, while the
majority are born as magnetars - neutron stars with huge surface magnetic
fields, ${\rm B\sim 10^{15}~Gauss}$ (Duncan and Thompson 1992; Thompson
and Duncan 1993, 1995, 1996). Their enormous magnetic field provides a
natural mechanism for braking these pulsars and spinning them down very
quickly. Such pulsars become unobservable in the radio in a very short
time due to their slow rotation and/or the fast decay of their magnetic
field. Their suggestion was motivated by the claimed success of the
magnetar model for soft gamma ray repeaters (SGRs) (e.g., Kouveliotou et
al. 1998; Kouveliotou et al. 1999) and anomalous X-ray pulsars (e.g.,
Gotthelf et al. 1999): The 4 known SGRs and the 6 known AXPs are all
slowly rotating (P=5-12 s) pulsars, surprisingly young, as evident from
the fact that all the 4 known SGRs and 3 out of the 6 known AXPs are found
in/near young SNRs (Cline et al.  1982; Kulkarni and Frail 1993; Vasisht
et al. 1994; Gotthelf \& Vasisht 1997:  Hurley et al. 1999b,c; Parmar et
al. 1998; Gaensler et al. 1999).  Their rotational energy is too small to
power either their persistent X-ray emission or the repeated gamma ray
bursts from SGRs. They spin down very quickly at a rate ${\rm \dot P\sim
10^{-10}-10^{-12}}$, too large to be explained by magnetic braking due to
${\rm B\leq 10^{13}}$ Gauss surface magnetic fields of ordinary radio
pulsars.  The huge magnetic fields of magnetars can explain both the fast
spin-down rate and the power source of SGRs and AXPs. 

However, the magnetar model for SGRs faces severe problems (Dar 1999; Dar
and DeR\'ujula 1999a). It faces an `age problem' - the magnetic spin-down
ages of SGRs, ${\rm\tau_s\leq P/2\dot P}$, are much shorter than the ages
of the expanding SNRs from their birth.  It also faces a `separation
problem' - the SGRs are found at distances too far from the centers of the
SNRs where they were born, if their magnetic spin down age is their true
age and their natal kick velocity is like that of ordinary pulsars, ${\rm
v_\perp\sim 350\pm 70~km~s^{-1}}$ (e.g., Lyne and Lorimer 1994; Toscano et
al 1999). Moreover, the magnetar model of SGRs and AXPs also faces an
`energy crisis':  Magnetic braking implies that their spin-down rate is
proportional to the square of their surface magnetic field.  If the energy
source of SGRs is their magnetic field energy, ${\rm E_m \approx B^2R^3/12}$,
then increase in their spin-down rate implies similar increase in their
magnetic field energy. Consequently, the large increase (more than
doubling) in the spin-down rate of SGR 1900+14 during its intensive burst
activity around August 27, 1998 (e.g., Marsden et al. 1999; Woods et al.
1999) implies a similar increase in its huge magnetic field energy
without a plausible source. Furthermore, although the surface magnetic
field of AXP 1E 2259+586 that was inferred from its observed
spin-down rate yields magnetic field energy much larger than its
rotational energy, it is not enough to power its steady X-ray luminosity ,
${\rm L_X\approx 8\times 10^{34}~erg ~s^{-1}}$, over its characteristic
age, ${\rm \tau_s\sim 1.5\times 10^5~y}$. Its inferred large magnetic
field (Thompson \& Duncan 1993) is also inconsistent with the absorption
features observed by ASCA in its X-ray spectrum (Corbet et al. 1995), if
they are interpreted as cyclotron lines. 

In view of the above difficulties of the magnetar model of SGRs and AXPs,
here we show that an alternative solution to the SGR and AXP puzzles
suggested by Dar (1999) and  Dar and DeR\'ujula (1999a) can also explain the 
dearth of young pulsars in SNRs and in the Galaxy.  We show that young 
Crab-like pulsars that are born in Type
II/Ib supernova explosions as fast rotating NSs with ordinary magnetic
fields, slow down quickly by relativistic particle emission along their
magnetic axis. When they have spun down sufficiently, they undergo a phase
transition and collapse to strange stars (SSs) or quark stars (QS) which
appear as SGRs and AXPs. The ejecta from the explosion, removes a
substantial fraction of their initial angular momentum. The gravitational
energy release, while they are cooling and contracting, powers both their
quiescent X-ray emission and their star quakes which produce `soft' gamma
ray bursts (Ramaty et al. 1980; Cheng et al. 1996). Relativistic particle
emission (RPE) in the form of jets along their magnetic axis 
spins them down very quickly into
slowly rotating, radio-quiet, dim X-ray pulsars or stellar black holes
while they are still in/near the SNR. Sensitive X-ray searches may find
these slowly rotating radio-quiet, dim X-ray pulsars or black holes
in/near many young SNRs. 
 
\section{Spin Down of Fast Rotating Pulsars} 

A spinning NS with a radius R, an angular velocity ${\rm \Omega=2\pi/P}$,
and a magnetic moment misaligned from the spin axis by an angle $\alpha$,
radiates electromagnetic energy at a rate (see, e.g., Shapiro and
Teukolsky 1983),
\begin{equation}
{\rm \dot E = {B_p^2 R^6\Omega^4 sin^2\alpha\over 6c^3}},
\end{equation}
where ${\rm B_p}$ is the magnetic field strength at the magnetic poles. 
Thus, braking the rotation of SGRs by magnetic dipole radiation requires 
a surface magnetic field 
\begin{equation}
{\rm B_p=\left({6c^3I\over R^6sin^2\alpha}\right)^{1/2}\sqrt{P\dot P}},
\end{equation}
where I is their moment of inertia. For a neutron star 
with a canonical mass of ${\rm 1.4M_\odot}$ which is supported
by the pressure of a non-relativistic degenerate Fermi gas of 
neutrons, ${\rm I\approx 10^{45}~gm~cm^{2}}$. Consequently,  
if the Crab pulsar has the canonical I and if it is slowing down by magnetic 
dipole radiation, then ${\rm B_p\approx 8\times 10^{12}/sin\alpha~Gauss}$
and its spin-down age is ${\rm \leq \tau_s= P/2\dot P\approx 1270~y}$,
which is in good agreement with its birth date, 1054 AD. This agreement
has been  considered as compelling evidence for magnetic braking
of Crab-like pulsars.    

Because of the $\Omega^4$ dependence, magnetic braking is a very
efficient spin-down mechanism for newly born NSs. However, 
relativistic particles emission (RPE) can also spin down efficiently 
fast rotating pulsars: The dipole magnetic field of pulsars drops with 
distance like ${\rm r^{-3}}$.  The relativistic particles which escape 
out from the pulsar's magnetic poles along the open magnetic 
lines co-rotate with the magnetic field until they reach a distance 
\begin{equation}
{\rm r_e \approx \left({cB^2R^6\over 2L_p}\right)^{1/4}}, 
\end{equation}
where their pressure becomes comparable to the magnetic pressure. 
But, for fast rotating pulsars 
with normal magnetic fields, 
${\rm r_e sin\alpha}$ becomes larger than the radius of the light 
cylinder ${\rm r_c= c/\Omega}$ (e.g., for  particle luminosity  
${\rm L_p\approx 10^{37}~erg}$, ${\rm r_e>r_c}$ 
for ${\rm P<100~ms}$). In that case, relativistic particles
of mass m, Lorentz factor $\gamma$ and energy ${\rm \epsilon=\gamma mc^2}$  
that stop co-rotating with the pulsar at the light cylinder carry an 
angular momentum $l={\rm =\gamma m c r_c=\epsilon/\Omega}$. The rate of 
angular momentum 
loss by emission of relativistic particles is then ${\rm \dot L\approx 
L_p/\Omega} $. Since ${\rm E_r=I\Omega^2/2}$ and ${\rm L=I\Omega}$,
one obtains that ${\rm\dot E_r=I\dot\Omega
\Omega=L\Omega}$, i.e.,  
\begin{equation}
{\rm \dot E_r=L_p},
\end{equation} 
and the spin-down time of the pulsar becomes ${\rm t_s\approx
E_r/L_p}$. 

Relation (4) is well satisfied, for instance, by the
Crab-pulsar where ${\rm \dot E_r=I\Omega\dot\Omega\approx 5\times
10^{38}~erg~s^{-1}}$, provided that the power input into the Crab
nebula ${\rm \approx 5\times 10^{38}~erg~s^{-1}}$  
(Manchester and Taylor 1977), is supplied by RPE from the
Crab-pulsar. This suggests that perhaps the surface magnetic field of the
Crab-pulsar (and of other young pulsars) which was inferred from its total
spin-down rate have been overestimated.  If most of its life the Crab
pulsar was spun down by RPE with 
${\rm L_p\approx 5\times 10^{38}~erg~s^{-1}}$ that is inferred from
the Crab nebula, and not by  magnetic dipole radiation,
then it was born with ${\rm P\approx 25~ms}$ and
was spun down to its present ${\rm P\approx 33~ms}$ during the past 945
years. The 25 ms period coincides approximately with the observed dividing
line between  ordinary isolated pulsars and `millisecond pulsars' which
are spun up by mass accretion in binaries (e.g., Toscano et al. 1998).

\section{Spin Down of SGRs and AXPs by RPE}

The non thermal quiescent X-ray emission and the highly suggestive radio
images (Vasisht et al. 1995; Frail et al. 1997) of SGR 1806-20 provide
compelling evidence for a steady RPE (relativistic jets ?) from the SGR. A
fading radio source within the localization of SGR 1900+14 has also been
interpreted as a short lived nebula powered by relativistic particles
ejected during the intense high energy activity of SGR 1900+14 in late
August 1998 (Frail et al. 1999). The emission of relativistic particles
along the magnetic axis can be a very efficient mechanism for braking
slowly rotating pulsars with normal magnetic fields, for which magnetic
braking is totally inefficientint (for approximate treatment of pulsar
braking by wind emission see, e.g., Harding et al. 1999 and references
therein). Exact magneto-hydrodynamic calculations of pulsar braking by
emission of relativistic particles, although desireable, is a formidable
project.  For our purpose, however, an approximate estimate is adequate: 

The angular momentum carried by a relativistic particle
that escapes the magnetic field at ${\rm
r_e<r_c}$ is, $l {\rm = \gamma m \Omega r_e^2 sin^2\alpha}$.
Hence, the rotational energy loss rate due to RPE 
with luminosity ${\rm L_p}$ is, 
\begin{equation} 
{\rm \dot E_r=I\Omega\dot\Omega \approx - B_p R^3
(L_p/2c^3)^{1/2} \Omega^2 sin^2\alpha}.
\end{equation}
It yields an exponential
decline, ${\rm E_r(t)=E_r(0)exp(-t/\tau_s)}$, with a characteristic time
\begin{equation} 
{\rm \tau_s={P\over 2\dot P}={I\over B_pR^3 sin^2\alpha}\left({c^3\over 
2L_p}\right)^{1/2}}. 
\end{equation}
For typical QS parameters (${\rm M\approx 1.4M_\odot}$, ${\rm R\approx
6~km}$, ${\rm B_pR^2\approx 10^{25}~Gauss~cm^2}$) and ${\rm L_p=10^{37}
~erg~s^{-1}}$, one obtains a short characteristic spin-down time, ${\rm 
\tau_s\sim 2000~y}$, for  GRBs and AXPs.  

Spin-down by RPE implies an increase in the spin-down rate of SGRs during
their gamma-ray bursts and radio flares which are caused, presumably, by
bursts of relativistic particles and return to their quiescent spin-down
rate at the end of the bursting activity. Indeed, the spin-down rate of
SGR 1900+14 increased dramatically during its intensive burst activity in
August 1998 (Woods et al. 1999 and references therein). After this
intensive activity it returned to its `quiescent' long term value prior to
the start of its intensive activity as shown in Fig 1.

\section{The Energy Source of SGRs and AXPs}

The energy source of SGRs and AXPs can be gravitational contraction
induced by cooling, spin-down and/or phase transition in pulsars (Dar
1999).  A cooling and spinning-down NS may undergo a series of phase
transitions (see, e.g., Shapiro and Teukolsky 1983 and references
therein): Cooper pairing of neutrons in cool NS can release a fraction of
MeV per pair, i.e., up to $\sim 10^{51}$ erg in NS. The NS can also undergo a
phase transition to a strange star (SS) and cooper pairing of quarks in NS
or SS can induce a phase transition and gravitational collapse to QS (Dar
1999; Dar and DeR\'ujula 1999a).  The contraction rate needed to power a 
total luminosity L is given by
\begin{equation} 
{\rm L\approx {1\over 2}{3(\Gamma-1)\over (5\Gamma-6)}\left({GM^2\over
R}\right) \left({\dot R\over R}\right)}, 
\end{equation} 
where $\Gamma=5/3$ for a pulsar which is supported mainly by Fermi
pressure of non-relativistic degenerate fermions and the factor 1/2 is
because $\sim 1/2$ of the gravitational binding 
energy is converted to Fermi motion. 
For a canonical 
pulsar mass ${\rm 1.4M_\odot}$ and a radius R=10 km, a contraction rate 
of e.g., ${\rm \dot R \sim 15~\mu m~ y^{-1}}$ can power 
persistent luminosity of ${\rm \sim 10^{37}~erg~s^{-1}}$.

Although quantum chromodynamics cannot provide yet reliable 
equation of state of quark matter at supernuclear densities, phase 
transitions to strange matter and quark matter are suggested by general 
considerations: The pressure of
cold nuclear matter at supernuclear densities is well approximated by that 
of degenerate Fermi gas of nucleons. The radius and central 
density of a self gravitating non-relativistic Fermi gas of neutrons of 
total baryonic mass M and zero angular momentum 
are given by the polytropic Emden-lane solution of the hydrostatic equation: 
\begin{equation}
{\rm R=\left({2\over 8^5\pi^4}\right)^{1/3}{h^2\over Gm_n^2}
       \left({M\over m_n}\right)^{-1/3}\approx 11.4
       \left({M\over M_\odot}\right)^{-1/3}~km}, 
\end{equation}
\begin{equation}
{\rm \rho_c\approx 6\bar\rho={9M\over 2\pi R^3}\approx 1.93\times 10^{15}
             \left({M\over M_\odot}\right)^{1/3}~g~cm^{-3}}. 
\end{equation}
However, the strangeness changing charge current reaction ${\rm
ud\rightarrow su}$ starts to transform neutrons at the top of the Fermi sea
into $\Lambda$'s at the bottom of the sea when the neutron Fermi energy
${\rm E_F=(h^2/8m_n)(3\rho/\pi m_n)^{2/3}}$
exceeds the effective ${\rm n\Lambda}$ mass difference,  
${\rm  c^2(m_{_\Lambda}-m_n)(1-GM(r)/c^2r)}$
where M(r) is the total mass enclosed within a radius r. For NS with zero 
angular
momentum it is satisfied for ${\rm M>1.27M_\odot}$. When the conversion ${\rm
n\rightarrow \Lambda}$  begins it reduces the Fermi pressure
and causes gravitational contraction which accelerates the conversion and
results in a gravitational collapse. The ${\rm n\rightarrow \Lambda}$ 
conversion stops when their chemical potentials equalize, i.e.,
when ${\rm E_F(n)-E_F(\Lambda)= c^2(m_{_\Lambda}-m_n)(1-GM(r)/c^2r)}$.

Because of their larger total number and smaller masses the total energy
and pressure of asymptotically-free degenerate relativistic Fermi gas of
quarks are higher than when they form bound nucleons. Consequently, most
probably, quarks are not liberated in the dense cores of neutron stars and
form there asymptotically-free degenerate Fermi gas. However, when cold
nuclear matter is compressed to high nuclear densities, it may convert
into a superfluid/superconducting Bose-condensate of spin zero diquarks
with antisymmetric color wave function that is several times as dense
(Alford et al.  1998;  Rapp et al. 1998; Wilczek 1998; Berges and
Rajagopal 1998).  Cooper pairing of quarks reduces the Fermi pressure in
the star and triggers its collapse which stops when the squeezing of the
Cooper pairs increases their internal energy by more than their pairing
energy (${\rm \sim 100~MeV}$).  The gravitational collapse to a strange or
a diquark star results in a mildly relativistic supernova-like explosion
and, possibly, in the ejection of highly relativistic jets.  The slow
gravitational contraction of the remnant QSs can power SGRs and AXPs (Dar
1999). The jets from the birth of QSs which happen to point in our
direction may produce the short duration gamma ray bursts (GRBs) (Dar
1999).  The long duration GRBs may be produced by highly relativistic jets
from the birth of black holes in gravitational collapse of NSs due to mass
accretion onto the proto-neutron star in supernovae explosions, or onto NS
in compact binary systems ( Dar and Plaga 1999, Dar and DeR\'ujula,
1999b).

\section{Observational Evidence For NS Collapse to QS?}
Since quantum chromodynamics cannot provide yet a definite answer to
whether spun-down pulsars undergo gravitational collapse, one must rely on
observations. Indeed, in addition of being able to explain the origin of
SGRs and AXPs, the gravitational collapse of spun-down pulsars to strange
or quark stars offers explanations for some other puzzling 
observations related to pulsars and SNRs: 

The steep spectra of the quiescent X-ray emission from SGRs and AXPs have
been interpreted as the characteristic Wein tail of a thermal black body
radiation from their surface. The Stefan-Boltzman law, ${\rm L_X=4\pi
R^2\sigma T^4\approx 1.3(R/10~km)^2(T/keV)^4\times 10^{37}~erg~s^{-1}}$
yields effective NS surface area, ${\rm A=4\pi R^2}$, significantly
smaller than that of NSs, ${\rm A_{NS}\approx 4\pi\times 10^2~km^2}$, as
summarized in Table I (provided that their measured thermal X-ray
emission and their distances are correct, and neglecting general
relativity effects).  An effective area smaller than the surface area of
NS may be due to a non uniform surface temperature. But, it can also be
due to the smaller radii of SSs or QSs. 

Radio observations expose a vast range of SNR shapes (e.g., Whiteoak\&
Green 1966). While very young SNRs have an expanding geometry, most older
SNRs have a distorted and complicated appearance. Their distortion has been
attributed to their expansion into an inhomogeneous interstellar medium
(ISM). However, some SNRs have striking symmetry properties which require
another explanation (e.g., Manchester 1987; Gaensler et al. 1998).  A second
explosion inside the SNR which also ejects relativistic jets along the
rotation axis, can explain the puzzling morphology of many SNRs (Dar and
DeR\'ujula 1999b). 

Imbalance in the momenta of oppositely ejected relativistic jets from the
collapse can impart to the SSs or QSs a natal kick that may explain the
observed large sky velocities of old, slowly rotating pulsars, whereas
millisecond pulsars and very young pulsars, which have high angular
momentum that prevents their collapse, have much smaller sky velocities
(Toscano et al.  1998 and references there in). (The small velocity of
millisecond pulsars may be a selection effect - only pulsars with a small
natal kick velocity can remain bound in a stellar binary and spun up by 
mass accretion).

\section{Discussion and Conclusions} 

RPE  can spin down very efficiently pulsars with
normal magnetic fields.  The slowly rotating pulsars are
probably much younger than inferred from magnetic braking. 

Various observations suggest that SGRs and AXPs are not magnetars, but,
most probably, they are young magnetized SSs or QSs formed recently by
gravitation collapse of NSs due to a phase transition in Crab-like pulsars
which have slowed-down sufficiently. The ejecta from the collapse can
remove a substantial fraction of the angular momentum of the NSs leaving
slowly rotating QSs. After their birth, gravitational contraction powers
their quiescent emission and their star quakes that produce their bursting
activity while they cool by radiation and spin-down very quickly by the
emission of relativistic particles along their magnetic axis. After a
short bursting activity, the SGRs, probably, become AXPs.  When they spin
down further and cool they may collapse to black holes or become
radio-quiet, dim X-ray sources.  Allowing for beaming and taking into
consideration that SNeII produce also black holes and millisecond pulsars
that are spun up in binaries, the observed number of SGRs and AXPs and
their spin-down times are consistent with their being a short, early stage
in the life of most of the ordinary pulsars.  Their quiescent thermal
X-ray emission suggests that they are much more condensed than NSs,
however, more precise measurements are needed to confirm that. 

Evidence for gravitational collapse of spun-down NSs to QSs or black holes
may come from double bang morphologies of SNRs, from measurements of
pulsar or black hole velocities in SNRs which do not point back to the
center of their SNRs, and perhaps from observations of gamma ray bursts
and SNe. When the SGRs/AXPs cool and spin down they become slowly
rotating, radio-quiet, X-ray dim pulsars or collapse to black holes. Most
of these dim pulsars or black holes, probably, are still present near/in
their SNRs.  Sensitive X-ray searches are required to establish their
presence there. 

Because most of the ordinary pulsars spin down in a relatively short time and
become radio-quiet pulsars, the number of ordinary pulsars that have been
detected in the Galaxy ($\sim 1000$) is smaller than that estimated from
their magnetic braking ages and sky velocities, by almost two orders of
magnitude. Millisecond pulsars have, relatively, very small magnetic fields.
Consequently, both magnetic braking and RPE cannot spin-down significantly
millisecond pulsars on time scales much smaller than the Hubble time. 
Therefore, the sky velocities of the vast majority of ordinary radio
pulsars point away from the Galactic plane while those of millisecond
pulsars have already been randomized by the Galactic gravitational fields
and are isotropic (e.g. Toscano et al. 1998).  Sensitive X-ray searches
may establish the presence of many relatively young, radio-quiet dim x-ray
pulsars in the Galaxy.

\noindent
{\bf Acknowledgements:} Helpful discussions, suggestions and comments   
by Alvaro DeR\'ujula and the kind  assistance of Shlomo 
Dado are gratefully acknowledged.

\footnotesize
\begin{deluxetable}{lcccccccc}
\tablecaption{Anomalous X-Ray Pulsars\vfill}
\label{tbl-1}
\tablehead{
\colhead{Pulsar} & \colhead{Ref.} & \colhead{P} &
\colhead{${\rm \tau_s}$} & 
\colhead{kT} & 
\colhead{${\rm L_X}~^a$}&\colhead{${\rm A_s}$}  \nl
\colhead{}  & \colhead{}     & \colhead{(s)}
& \colhead{(yrs)} & \colhead{(keV)} 
& \colhead{(${\rm 10^{35}~erg~s^{-1}}$)}
& \colhead{(${\rm 4\pi\times 10^2 km^2}$)} }
\startdata
1E 1841$-$045 & b & 11.76 & $3.9\times 10^3 $&0.55 
& $3 d_{7 kpc}^2$ & $0.25d_{7kpc}^2$  \nl
1E 2259$+$586  & c & 6.98  & $1.5 \times 10^{5}$&0.41
& $0.8 d_{4 kpc}^2$ & $ 0.22d_{4 kpc}^2$   \nl
4U 0142$+$615 & d & 8.69  & $6.0\times 10^4$ & 0.39
& $0.7 d_{4 kpc}^2$ & $ 0.24d_{1 kpc}^2$ \nl
1E 1048$-$5937& e & 6.44  & $4.6\times 10^{3}$ & 0.64 
& $5 d_{10 kpc}^2$ & $ 0.23d_{10 kpc}^2$ & \nl
RX J170849.0$-$400910& f & 11.00 & \dots & 0.40
& $10 d_{10 kpc}^2$ & $0.17d_{10kpc}^2$ \nl
PSR J1844-0258 & g & 6.97  &    \dots &0.64
& $3 d_{15 kpc}^2$ & $ 0.15 d_{15 kpc}^2$ \nl
\enddata
\noindent
\tablenotetext{a}{All luminosities are are in the $\sim 1-10$ keV energy band
as corrected by Gotthelf and Vasisht 1998 for absorption.\\
$^b$ Vasisht \& Gotthelf 1997.\\ 
$^c$ Corbet et al. 1995, Iwasawa et al. 1992, and refs. therein.\\
$^d$ White et al. 1996, Mereghetti \& Stella 1995, and refs. therein.\\
$^e$ Parmar et al. 1998, Mereghetti et al. 1997, Corbet \& Mihara
1997 and refs. therein.\\
$^f$ Sugizaki et al. 1997.\\
$^g$ Gaensler et al. 1999.}
\end{deluxetable}

\newpage 
\begin{figure} 
\plotone{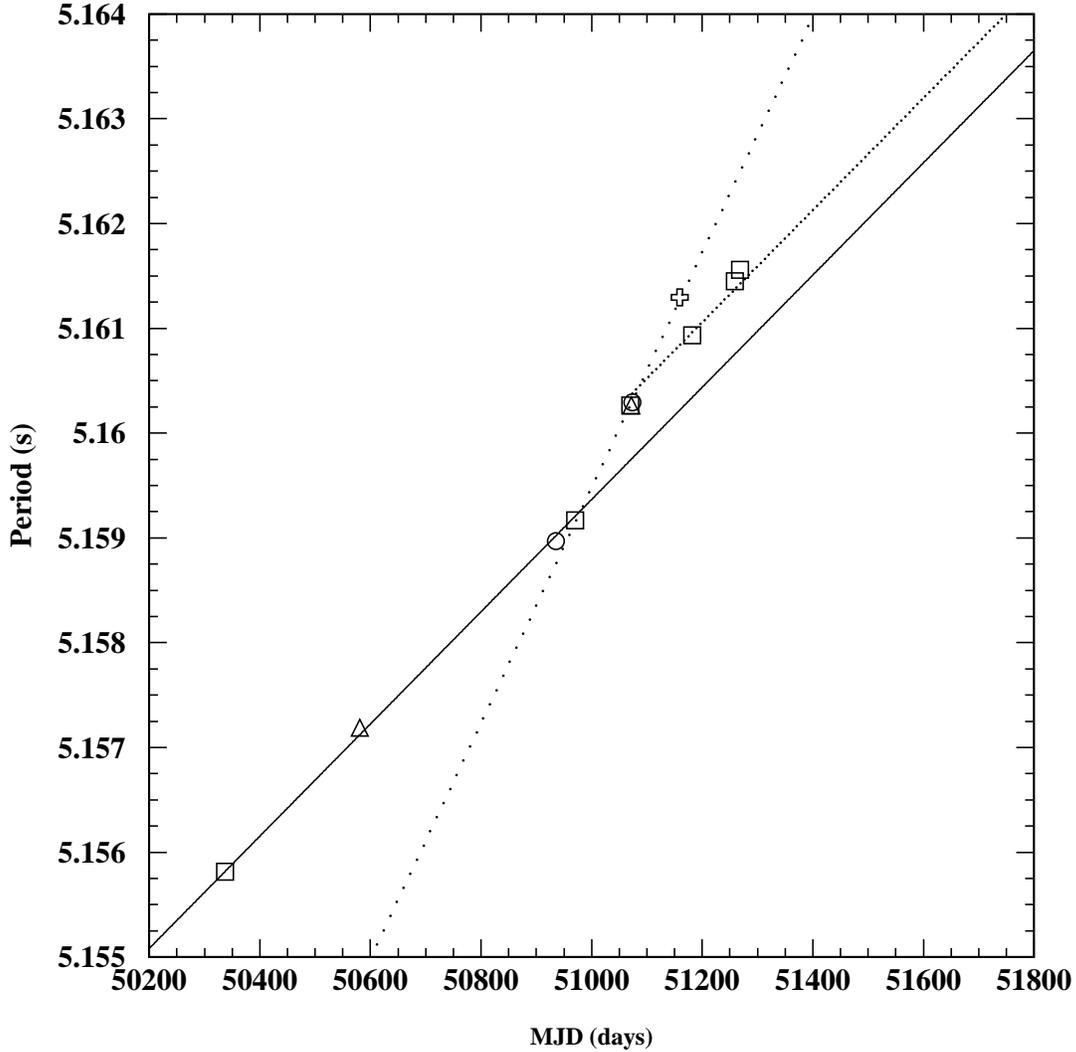} 
\caption{The period of SGR
1900+14 as function of time measured by RXTE (Woods et al. 1999,
squares), BeppoSAX (Woods et al. 1999, triangles), ASCA (Hurley et al.
1999c; Murakami et al 1999, circles) and BSA (Shitov 1999, cross). The
lines are best linear fits to the X-ray periods of the SGR before June
9, 1998 (${\rm \dot P=6\times 10^{-11}}$), between June 9 - August 27,
1998 (${\rm \dot P=1.3\times 10^{-10}}$), and after August 27, 1998 (${\rm
\dot P=6\times 10^{-11}}$).  Between June 9-August 28, its `average'
spin-down rate has changed by a factor $\sim 2.2$ as a result of
continuous spin-down or a sudden brake}. 
\end{figure}

\begin{references}



\reference{}  
Alford, M. et al. 1998, Phys. Lett. B422, 247    
\reference{}
Berges, J. \& Rajagopal, 1999, Nucl. Phys. B538, 215
\reference{}
Camilo, F. et al. astro-ph/9911185
\reference{}
Cheng et al. 1996, Nature 382, 518
\reference{}
Cline, T. L. et al. 1982, ApJ,  255, L45
\reference{}
Corbet, R. H. D., et al. 1995, ApJ, 433, 786
\reference{} 
Corbet, R. H. D. \& Mihara 1997, ApJ 475, L127
\reference{}
Dar, A. 1999, A\&A  Suppl. Ser. 138 (3), 505 
\reference{}
Dar, A. and DeR\'ujula, A. 1999a, submitted for publication 
\reference{}
Dar, A. and DeR\'ujula, A. 1999b, in preparation 
\reference{}
Dar, A. \& Plaga, R. 1999, A\&A 349, 259 
\reference{}  
Duncan, R. C. \& Thompson, C. 1992, ApJ, 392, L9 
\reference{} 
Frail, D. A. et al. 1997, ApJ, 480,  L129 
\reference{} 
Frail, D. A. et al. 1999, Nature, 398, 127 
\reference{} 
Gaensler, B. M.  et al. 1998, MNRAS 299, 812
\reference{} 
Gaensler, B. M.  et al. 1998, ApJ, 526, L37
\reference{} 
Gaensler, B. M. and Johnston, S. 1995, MNRAS, 277, 1243
\reference{} 
Gotthelf, E. V. \& Vasisht, G. 1997,  ApJ, 486, L133
\reference{} 
Gotthelf, E. V. \& Vasisht, G. 1998,  astro-ph/9804025
\reference{} 
Gotthelf, E. V. \& Vasisht, G. 1999,  astro-ph/9909139
\reference{} 
Gotthelf, E. G. et al. 1999, ApJ, 522, L49
\reference{}
Green, D. A. 1998 http://www.mrao.cam.ac.uk/surveys/snrs 
\reference{}
Harding, A. et al. 1999, ApJ, 525, L125  
\reference{}
Hurley, K.  et al. 1999a, ApJ, 510, L111
\reference{}
Hurley, K.  et al. 1999b, ApJ, 519, L143
\reference{}
Hurley, K.  et al. 1999c, ApJ, 523, L37
\reference{} 
Iwasawa, K. et al. 1992, PASJ, 44, 9
\reference{}
Kouveliotou, C. et al. 1998, Nature, 393, 235
\reference{}
Kouveliotou, C. et al. 1999, ApJ, 510, L115
\reference{} 
Kulkarni, S. R. \& Frail, D. A. 1993, Nature, 365, 33   
\reference{}
Lyne, A. G. \& Lorimer, D. R. 1994, Nature, 369, 127
\reference{} 
Manchester, R. N. \&  Taylor, J. H. 1977, {\it Pulsars}, Freeman, San
Francisco, California.   
\reference{} 
Manchester, R. N. 1987, A\&A 171, 205
\reference{} 
Marsden, D. et al. 1999, ApJ 520 L107
\reference{} 
Mazets, E. et al. 1979a, Nature 282, 587
\reference{} 
Mazets, E. et al. 1979b, Sov. Astro. Lett. 5, 343
\reference{} 
Mereghetti, S. astro-ph/9911252
\reference{} 
Mereghetti, S. \& Stella, L. 1995, ApJ, 442, L17
\reference{} 
Mereghetti, S. et al. 1997, A\&A, 321, 835
\reference{}
Parmar, A. et al. 1998, A\&A, 330, 175
\reference{}
Ramaty, R. et al., 1980, Nature, 287, 122
\reference{}
Rapp, R. et al. 1998, PRL, 81, 53  
\reference{} 
S. L. Shapiro and S. A. Teukolsky, 1983, Black Holes, White Dwarfs and 
Neutron Stars, John Wiley \& Sons    
\reference{} 
Sugizaki et al. 1997, PASJ, 49, L25
\reference{} 
Thompson, C. \& Duncan, R. C. 1993, ApJ, 408, 194
\reference{} 
Thompson, C. \& Duncan, R. C. 1995, MNRAS, 275, 255
\reference{} 
Thompson, C. \& Duncan, R. C. 1996, ApJ, 473, 332
\reference{}
Toscano, M. et al. 1999, MNRAS 307, 925
\reference{}
van den Bergh, S. \& Tamman, G.A. 1991  ARA\&A 29, 363 
\reference{} 
Vasisht, G. et al. 1994, ApJ, 431, L35
\reference{} 
Vasisht, G. et al. 1995, ApJ, 440, L65  
\reference{} 
Vasisht, G. \& Gotthelf, E. V. 1997, ApJ, 486, L129
\reference{} 
White, N. E. et al. 1996, ApJ, 463 L83
\reference{}
Whiteoak, J. B. Z. \& Green, A. J., 1996, A\&AS 118, 329 
\reference{}
Wilczek, F. 1998, Nature  395, 220  
\reference{}
Woods, P. M. et al. 1999b, ApJ, 524, L55   


\end{references}
\end{document}